\documentclass[onecolumn,showpacs,preprintnumbers,amsmath,amssymb,eqsecnum]{revtex4}
\usepackage{dcolumn}
\usepackage{bm}
\usepackage[dvips]{graphicx}
\usepackage{wrapfig}
\usepackage{psfrag}
\usepackage{color}
\begin{document}

\def\sh{\mathop{\rm sh}\nolimits}
\def\ch{\mathop{\rm ch}\nolimits}
\def\var{\mathop{\rm var}}\def\exp{\mathop{\rm exp}\nolimits}
\def\Re{\mathop{\rm Re}\nolimits}
\def\Sp{\mathop{\rm Sp}\nolimits}
\def\kp{\mathop{\text{\ae}}\nolimits}
\def\bk{{\bf {k}}}
\def\bp{{\bf {p}}}
\def\bq{{\bf {q}}}
\def\lra{\mathop{\longrightarrow}}
\def\Const{\mathop{\rm Const}\nolimits}
\def\sh{\mathop{\rm sh}\nolimits}
\def\ch{\mathop{\rm ch}\nolimits}
\def\var{\mathop{\rm var}}
\def\mK{\mathop{{\mathfrak {K}}}\nolimits}
\def\mR{\mathop{{\mathfrak {R}}}\nolimits}
\def\mv{\mathop{{\mathfrak {v}}}\nolimits}
\def\mV{\mathop{{\mathfrak {V}}}\nolimits}
\def\mD{\mathop{{\mathfrak {D}}}\nolimits}
\def\mN{\mathop{{\mathfrak {N}}}\nolimits}
\def\mS{\mathop{{\mathfrak {S}}}\nolimits}

\newcommand\ve[1]{{\mathbf{#1}}}

\def\Re{\mbox {Re}}
\newcommand{\Z}{\mathbb{Z}}
\newcommand{\R}{\mathbb{R}}
\def\mK{\mathop{{\mathfrak {K}}}\nolimits}
\def\mL{\mathop{{\mathfrak {L}}}\nolimits}
\def\mk{\mathop{{\mathfrak {k}}}\nolimits}
\def\mR{\mathop{{\mathfrak {R}}}\nolimits}
\def\mv{\mathop{{\mathfrak {v}}}\nolimits}
\def\mV{\mathop{{\mathfrak {V}}}\nolimits}
\def\mD{\mathop{{\mathfrak {D}}}\nolimits}
\def\mN{\mathop{{\mathfrak {N}}}\nolimits}
\def\ml{\mathop{{\mathfrak {l}}}\nolimits}
\def\mf{\mathop{{\mathfrak {f}}}\nolimits}
\def\me{\mathop{{\mathfrak {e}}}\nolimits}
\newcommand{\ccm}{{\cal M}}
\newcommand{\cE}{{\cal E}}
\newcommand{\cV}{{\cal V}}
\newcommand{\cI}{{\cal I}}
\newcommand{\cR}{{\cal R}}
\newcommand{\cK}{{\cal K}}
\newcommand{\cH}{{\cal H}}
\newcommand{\cW}{{\cal W}}
\newcommand{\cS}{{\cal S}}
\newcommand{\cT}{{\cal T}}
\newcommand{\ce}{{\cal e}}

\def\br{\mathop{{\bf {r}}}\nolimits}
\def\bS{\mathop{{\bf {S}}}\nolimits}
\def\bA{\mathop{{\bf {A}}}\nolimits}
\def\bJ{\mathop{{\bf {J}}}\nolimits}
\def\bn{\mathop{{\bf {n}}}\nolimits}
\def\bg{\mathop{{\bf {g}}}\nolimits}
\def\bv{\mathop{{\bf {v}}}\nolimits}
\def\be{\mathop{{\bf {e}}}\nolimits}
\def\bp{\mathop{{\bf {p}}}\nolimits}
\def\bz{\mathop{{\bf {z}}}\nolimits}
\def\bbf{\mathop{{\bf {f}}}\nolimits}
\def\bb{\mathop{{\bf {b}}}\nolimits}
\def\ba{\mathop{{\bf {a}}}\nolimits}
\def\bx{\mathop{{\bf {x}}}\nolimits}
\def\by{\mathop{{\bf {y}}}\nolimits}
\def\br{\mathop{{\bf {r}}}\nolimits}
\def\bs{\mathop{{\bf {s}}}\nolimits}
\def\bH{\mathop{{\bf {H}}}\nolimits}
\def\bk{\mathop{{\bf {k}}}\nolimits}
\def\be{\mathop{{\bf {e}}}\nolimits}
\def\bnul{\mathop{{\bf {0}}}\nolimits}
\def\bq{{\bf {q}}}

\newcommand{\oV}{\overline{V}}
\newcommand{\vkp}{\varkappa}
\newcommand{\os}{\overline{s}}
\newcommand{\opsi}{\overline{\psi}}
\newcommand{\ov}{\overline{v}}
\newcommand{\oW}{\overline{W}}
\newcommand{\oPhi}{\overline{\Phi}}

\def\mI{\mathop{{\mathfrak {I}}}\nolimits}
\def\mA{\mathop{{\mathfrak {A}}}\nolimits}

\def\diag{\mathop{\rm diag}\nolimits}
\def\st{\mathop{\rm st}\nolimits}
\def\tr{\mathop{\rm tr}\nolimits}
\def\sign{\mathop{\rm sign}\nolimits}
\def\d{\mathop{\rm d}\nolimits}
\def\const{\mathop{\rm const}\nolimits}
\def\O{\mathop{\rm O}\nolimits}
\def\Spin{\mathop{\rm Spin}\nolimits}
\def\exp{\mathop{\rm exp}\nolimits}
\def\SU{\mathop{\rm SU}\nolimits}
\def\mU{\mathop{{\mathfrak {U}}}\nolimits}
\newcommand{\cU}{{\cal U}}
\newcommand{\cD}{{\cal D}}

\def\mC{\mathop{{\mathfrak {C}}}\nolimits}
\def\mI{\mathop{{\mathfrak {I}}}\nolimits}
\def\mA{\mathop{{\mathfrak {A}}}\nolimits}
\def\mU{\mathop{{\mathfrak {U}}}\nolimits}

\def\st{\mathop{\rm st}\nolimits}
\def\tr{\mathop{\rm tr}\nolimits}
\def\sign{\mathop{\rm sign}\nolimits}
\def\d{\mathop{\rm d}\nolimits}
\def\const{\mathop{\rm const}\nolimits}
\def\O{\mathop{\rm O}\nolimits}
\def\Spin{\mathop{\rm Spin}\nolimits}
\def\exp{\mathop{\rm exp}\nolimits}

\title{Domain wall between the Dirac sea and the "anti-Dirac sea"}

\author {S.N. Vergeles\vspace*{4mm}\footnote{{e-mail:vergeles@itp.ac.ru}}}

\affiliation{Landau Institute for Theoretical Physics,
Russian Academy of Sciences,
Chernogolovka, Moscow region, 142432 Russia \linebreak
and   \linebreak
Moscow Institute of Physics and Technology, Department
of Theoretical Physics, Dolgoprudnyj, Moskow region,
141707 Russia}

\begin{abstract}
It was shown in work \cite{vergeles2021note} that in the theory of gravity coupled with the Dirac field, each state $|\lambda\rangle$ has its own twin $|\lambda;PT\rangle$, which is obtained by a discrete PT transformation. If in the state $|\lambda\rangle$ the Dirac sea is filled, then in the state $|\lambda;PT\rangle$ there is an "anti-Dirac" filling (in terms of the state $|\lambda\rangle$). It is important that the energies of these states are the same. Therefore, there may be domains with different filling of the Dirac sea. Here we study a domain wall connecting two such adjacent domains.
\end{abstract}


\maketitle

\section{Introduction}

We study some properties of the theory of gravity coupled with the Dirac field. The theory is assumed to be lattice-regularized. The version of the lattice theory of gravity, which is used here, is discussed in detail in \cite{vergeles2021note}. In the naive long-wavelength limit, this lattice theory transforms into the considered continuum theory. It is convenient to represent the action of the theory of gravity in the form of Palatini, when the gravitational degrees of freedom are described by a tetrad and a connection.

It was shown in \cite{vergeles2021note}  that in the lattice version of the theory (and thus in the long-wavelength limit) there is a global discrete symmetry, which is an analogue of the well-known $PT$-symmetry in field theory. We therefore call this discrete symmetry as $PT$-symmetry because Dirac fields are transformed at each vertex of the lattice (or point in space-time) according to the law of combined $P$ and $T$ transformation known in flat field theory. The difference between our proposed $PT$-transformation and the same in the flat field theory is that in the latter case the coordinates (which are global) change sign, while we propose to change the sign of the tetrad. In this case, the connection coefficients remain unchanged.
Such a rule for transforming variables in the case of $PT$-reflection is dictated by the lattice model of the theory. Indeed, there are no coordinates on the lattice at all, there are only dynamic variables assigned to the lattice elements. Therefore, the entire load of any transformation must be borne by dynamic variables.
To this it should be added that in the continual theory of gravity, the change in the sign of local coordinates seems to us not quite correct for the reason that different local coordinates mutually transform (generally speaking) according to a nonlinear law.

Since, as in the case of flat field theory, repeated application of the PT transformation gives the identical transformation, this means that there is $Z_2$-symmetry in the theory. This $Z_2$-transformation mutually replaces particles and antiparticles. To clarify the situation, it is useful to consider the case of Minkowski space, when it is correct to consider Dirac modes with a certain energy. As shown in (\ref{Dirac_Mode_Trans}), as a result of the $PT$-transformation, the mode energy changes sign. But one must take into account the fact that the tetrad also changes sign under the $PT$-transformation, and the Dirac Hamiltonian in curved space-time is a third-degree functional relative to the tetrad. Therefore, the energy of the transformed ground state (that is, the energy of the "anti-Dirac sea") remains negative, although from the point of view of an observer related with the initial ground state in Minkowski space-time, Dirac modes with positive energies in the transformed ground state are filled. Therefore, we call the transformed vacuum the "anti-Dirac sea".

To some extent, the described situation resembles the situation in the 2D Ising model
which also has $Z_2$-symmetry: this model is symmetric relative to the global reversal of the sign of all spins. At temperatures below critical the average spin value is not zero, but it can be positive or negative with equal Helmholtz free energy density. Therefore, in the Ising model, there can be adjacent domains with opposite sign values of the mean spin. A similar phenomenon takes place in the lattice theory of gravity considered here.
Here the slip of the tongue "in the lattice theory of gravity" is not accidental.
The point is that in any lattice field theory the energies or quasi-energies of quasiparticles are limited in absolute value. Therefore, in the process of the Big Bang of the Universe, domains can be formed, the states of which differ in the indicated $PT$-transformation.

The subject of the proposed work is the study of a domain wall.

The work is organized as follows:

In Section II, the necessary notations are introduced, the studied model is described both in the continuum and on the lattice, and the global $PT$-transformation is determined. Further, some technical means are presented that are used in the calculations in the following sections.

Section III contains several solutions for a flat domain wall. Only the solution for a domain wall moving at the speed of light seems to be physically acceptable.

Section IV contains some speculation about the distribution of fermionic charge between domains. Although the total fermionic charge of the entire system is strictly conserved on the lattice and it can be zero, in one of the domains its density can be positive, and in the other it can be negative. This raises a fundamental question: is gravity a "player" in the problem of the baryon asymmetry of the Universe?

\section{Description of the model under study, as well as the technical means used}

Consider the theory of gravity coupled with the Dirac field. The action of this system is as follows:
\begin{gather}
\mA=\mA_g+\mA_{\Psi}+\mA_{\Lambda},
\label{Act_tot}
\end{gather}
where
\begin{gather}
\mA_g=-\frac{1}{4\,l^2_P}\varepsilon_{abcd}\int\mR^{ab}\wedge e^c\wedge e^d=
-\frac{1}{4\,l^2_P}\varepsilon_{abcd}\varepsilon^{\mu\nu\lambda\rho}\int\mR^{ab}_{\mu\nu}e^c_{\lambda}
e^d_{\rho}\d x^0\wedge\d x^1\wedge\d x^2\wedge\d x^3,
\nonumber \\
\frac12\mR^{ab}=\d \omega^{ab}+\omega^a_c\wedge\omega^{cb}, \quad \omega^{ab}=\omega^{ab}_{\mu}\d x^{\mu},
\quad e^a=e^a_{\mu}\d x^{\mu},
\label{Long_Wav_Grav_Act}
\end{gather}
\begin{gather}
\mA_{\Psi}=\frac16\varepsilon_{abcd}\int\Theta^a\wedge e^b\wedge e^c\wedge e^d=
\frac16\varepsilon_{abcd}\varepsilon^{\mu\nu\lambda\rho}\int\Theta^a_{\mu}\,e^b_{\nu}\,e^c_{\lambda}\,e^d_{\rho}
\d x^0\wedge\d x^1\wedge\d x^2\wedge\d x^3,
\nonumber \\
\Theta^a=\frac{i}{2}\left[\overline{\Psi}\gamma^a{\cal D}_{\mu}\,\Psi-
\left(\overline{{\cal D}_{\mu}\,\Psi}\right)\gamma^a\Psi\right]\d x^{\mu}\equiv\Theta^a_{\mu}\d x^{\mu}, \quad
{\cal D}_{\mu}=\left(\partial_{\mu}+\frac12\sigma^{ab}\omega_{ab\mu}\right),
\label{Long_Wav_Dir_Act}
\end{gather}
\begin{gather}
\mA_{\Lambda}=\frac18\Lambda\varepsilon_{abcd}\int e^a\wedge e^b\wedge e^c\wedge e^d=
\frac18\Lambda\varepsilon_{abcd}\varepsilon^{\mu\nu\lambda\rho}\int e^a_{\mu}\,e^b_{\nu}\,e^c_{\lambda}\,e^d_{\rho}
\d x^0\wedge\d x^1\wedge\d x^2\wedge\d x^3.
\label{Cosm_Const_Act}
\end{gather}
Everywhere in the continuum theory there is a Minkowski signature, Levi-Civita symbols are equal to units if their indices are ordered as $(0123)$, the symbols $\d$ and $\wedge$ denote external differentiation and external multiplication of differential forms, respectively, $\omega^{ab}_{\mu}$ denote connection coefficients in some orthonormal basis. We also write down the expression for the metric tensor in terms of the tetrad:
$g_{\mu\nu}=\eta_{ab}e^a_{\mu}e^b_{\nu}$, $\eta_{ab}=\diag(1,-1,-1,-1)$. Everywhere small Latin and Greek indices run through the values 0,1,2,3.

In this paper, we assume a lattice regularization of the theory under consideration. The lattice is a 4D orientable simplicial complex $\mK$ on which the action of gravity coupled with the Dirac field is determined.
On the lattice, the signature is assumed to be Euclidean.
The reason the Euclidean signature is switched in lattice theories is typical of all gauge theories: fixing a gauge in lattice theories, especially on irregular lattices, is impossible. Therefore, for the integral over the gauge field to converge, the gauge group must be compact. The latter property is provided by the Euclidean signature.
For a detailed presentation and discussion of the theory of gravity on a lattice, see  \cite{vergeles2021note}.
There, the transfer from the Euclidean signature on a lattice to the Minkowski signature in the long-wavelength limit is also considered in detail.
However, for ease of reading, we provide some definitions here.

Suppose that any of its 4-simplexes belongs to such a finite (or infinite) sub-complex ${\mK}'\in\mK$  which has a geometric realization in  $\R^4$ topologically equivalent to a disk  without cavities.
The vertices are designated as
$a_{\cV}$, the indices ${\cV}=1,2,\dots,\,{\mN}^{(0)}\rightarrow\infty$ and ${\cW}$ enumerate the vertices
and 4-simplices, correspondingly. It is necessary to use
the local enumeration of the vertices $a_{\cV}$ attached to a given
4-simplex: the all five vertices of a 4-simplex with index ${\cW}$
are enumerated as $a_{{\cV}_{({\cW})i}}$, $i=1,2,3,4,5$. The later notations with extra low index  $({\cW})$
indicate that the corresponding quantities belong to the
4-simplex with index ${\cW}$. Of course, these same quantities  also belong to another 4-simplex with index  ${\cW}'$, and 4-simplexes with indices  ${\cW}$ and  ${\cW}'$ must be adjacent.
The Levi-Civita symbol with in pairs
different indexes
$\varepsilon_{{\cV}_{({\cW})1}{\cV}_{({\cW})2}{\cV}_{({\cW})3}{\cV}_{({\cW})4}{\cV}_{({\cW})5}}=\pm 1$ depending on
whether the order of vertices
$s^4_{\cW}=a_{{\cV}_{({\cW})1}}a_{{\cV}_{({\cW})2}}a_{{\cV}_{({\cW})3}}a_{{\cV}_{({\cW})4}}a_{{\cV}_{({\cW})5}}$ defines the
positive or negative orientation of 4-simplex $s^4_{\cW}$.
An element of the compact group $\Spin(4)$ and an element of the Clifford algebra
\begin{gather}
\Omega_{{\cV}_1{\cV}_2}=\Omega^{-1}_{{\cV}_2{\cV}_1}=
\exp\left(\omega_{{\cV}_1{\cV}_2}\right)
\in\Spin(4), \quad
\omega_{{\cV}_1{\cV}_2}\equiv\frac{1}{2}\sigma^{ab}
\omega^{ab}_{{\cV}_1{\cV}_2}=-\omega_{{\cV}_2{\cV}_1},
\nonumber \\
\hat{e}_{{\cV}_1{\cV}_2}\equiv
-\Omega_{{\cV}_1{\cV}_2}\hat{e}_{{\cV}_2{\cV}_1}\Omega_{{\cV}_1{\cV}_2}^{-1}
\equiv e^a_{{\cV}_1{\cV}_2}\gamma^a,  \quad
-\infty<e^a_{{\cV}_1{\cV}_2}<+\infty,
\nonumber
\end{gather}
are assigned for each oriented 1-simplex $a_{{\cV}_1}a_{{\cV}_2}$.

The conjecture is that the set of variables $\{\Omega,\,\hat{e}\}$  is an independent set of dynamic variables.
Fermionic degrees of freedom (Dirac spinors) are assigned to each vertex of the complex:
\begin{gather}
\Psi^{\dag}_{\cV}, \quad \Psi_{\cV}.
\nonumber
\end{gather}
The set of variables $\{\Psi^{\dag},\,\Psi\}$ is also a set of mutually independent variables, and the spinors $\Psi^{\dag}_{\cV}$ and $\Psi_{\cV}$ are in mutual involution (or anti-involution) relative to Hermitian conjugation operation.

Let's write out the corresponding action:
\begin{gather}
\mA_g=\frac{1}{5\cdot
24\cdot2\cdot l_P^2}\sum_{\cW}\sum_{\sigma}
\varepsilon_{\sigma}
\nonumber \\
\times\tr\gamma^5\bigg\{
\Omega_{\sigma({\cV}_{({\cW})5})\sigma({\cV}_{({\cW})1})}
\Omega_{\sigma({\cV}_{({\cW})1})\sigma({\cV}_{({\cW})2})}\Omega_{\sigma({\cV}_{({\cW})2})\sigma({\cV}_{({\cW})5})}
\hat{e}_{\sigma({\cV}_{({\cW})5})\sigma({\cV}_{({\cW})3})}
\hat{e}_{\sigma({\cV}_{({\cW})5})\sigma({\cV}_{({\cW})4})}\bigg\},
\nonumber \\
\hat{e}_{{\cV}_1{\cV}_2}=e^a_{{\cV}_1{\cV}_2}\gamma^a,  \quad
\Omega_{{\cV}_1{\cV}_2}=\Omega^{-1}_{{\cV}_2{\cV}_1}\in\Spin(4).
\label{Latt_Action_Grav}
\end{gather}
The index ${\cW}$ enumerates the 4-simplexes, the index ${\cV}_{\cW}$ enumerates the vertices belonging to the 4-simplex with the index
${\cW}$, $\sigma$ denotes one of the $5!$ permutations of the vertices of a given 4-simplex and $\varepsilon_{\sigma}=\pm1$ depending on the parity of the permutation $\sigma$.
\begin{gather}
\mA_{\Psi}=-\frac{1}{5\cdot24^2}\sum_{\cW}\sum_{\sigma}
\varepsilon_{\sigma}
\nonumber \\
\times\tr\gamma^5\bigg\{ \hat{\Theta}_{\sigma({\cV}_{({\cW})5})\sigma({\cV}_{({\cW})1})}
\hat{e}_{\sigma({\cV}_{({\cW})5})\sigma({\cV}_{({\cW})2})}
\hat{e}_{\sigma({\cV}_{({\cW})5})\sigma({\cV}_{({\cW})3})}
\hat{e}_{\sigma({\cV}_{({\cW})5})\sigma({\cV}_{({\cW})4})}\bigg\},
\label{Latt_Action_Ferm}
\end{gather}
\begin{gather}
\hat{\Theta}_{{\cV}_1{\cV}_2}\equiv
\frac{i}{2}\gamma^a\left(\Psi^{\dag}_{{\cV}_1}\gamma^a
\Omega_{{\cV}_1{\cV}_2}\Psi_{{\cV}_2}-\Psi^{\dag}_{{\cV}_2}\Omega_{{\cV}_2{\cV}_1}\gamma^a\Psi_{{\cV}_1}\right)
\equiv\Theta^a_{{\cV}_1{\cV}_2}\gamma^a=\hat{\Theta}_{{\cV}_1{\cV}_2}^{\dag},
\label{Dirac_Form}
\end{gather}
\begin{gather}
\mA_{\Lambda}=\frac{1}{32}\Lambda\sum_{\cW}
\sum_{\sigma}
\varepsilon_{\sigma}
\nonumber \\
\times\tr\bigg\{\gamma^5
\hat{e}_{\sigma({\cV}_{({\cW})5})\sigma({\cV}_{({\cW})1})}
\hat{e}_{\sigma({\cV}_{({\cW})5})\sigma({\cV}_{({\cW})2})}
\hat{e}_{\sigma({\cV}_{({\cW})5})\sigma({\cV}_{({\cW})3})}
\hat{e}_{\sigma({\cV}_{({\cW})5})\sigma({\cV}_{({\cW})4})}\bigg\}.
\label{Lat_Cosm_Const_Act}
\end{gather}
In the naive long-wavelength limit and as a result of the transition to the Minkowski signature, the action
(\ref{Latt_Action_Grav})-(\ref{Lat_Cosm_Const_Act}) goes into action
(\ref{Long_Wav_Grav_Act})-(\ref{Cosm_Const_Act}).

Actions (\ref{Act_tot})-(\ref{Cosm_Const_Act}) and (\ref{Latt_Action_Grav})-(\ref{Lat_Cosm_Const_Act}) are invariant under gauge transformations \cite{vergeles2021note}, which we are not interested in here. The same actions are invariant under the following global discrete transformations.

In the long-wavelength limit, in the case of the Minkowski signature, the action (\ref{Act_tot})-(\ref{Cosm_Const_Act})
is invariant under the following global discrete transformation of variables:
\begin{gather}
\left(\Psi^{PT}\right)(x)=U_{PT}\left(\overline{\Psi}(x)\right)^t, \quad
\left(\overline{\Psi}^{PT}\right)(x)=-\left(\Psi(x)\right)^tU^{-1}_{PT},
\quad  U^{-1}_{PT}\gamma^aU_{PT}=(\gamma^a)^t,
\nonumber \\
\left(e^{PT}\right)^a_{\mu}(x)=-e^{a}_{\mu}(x), \quad
\left(\omega^{PT}\right)^{ab}_{\mu}(x)=\omega^{ab}_{\mu}(x),
\nonumber \\
U_{PT}=i\gamma^3\gamma^1.
\label{PT_trans_Cont_Minkovski}
\end{gather}
Superscript ${}^t$ means matrix transposition. It follows from (\ref{PT_trans_Cont_Minkovski}) and (\ref{Long_Wav_Dir_Act}) that \cite{vergeles2021note}
\begin{gather}
\left(\Theta^{PT}\right)^a_{\mu}=-\Theta^a_{\mu}.
\label{PT_trans_Dir_Ac_Min}
\end{gather}
This transformation is analogous to the PT transformation. Indeed, the transformation of Dirac fields in (\ref{PT_trans_Cont_Minkovski}) exactly coincides with the combined transformation of $P$ and $T$ in continuous quantum field theory in Minkowski space. The difference with the generally accepted version of the PT transformation is that there is no reflection of the coordinates $x\longrightarrow-x$, but only the dynamic variables of the theory are transformed.
In the theory of gravity, this approach seems more natural, since the role of local coordinates can be played by any 4 independent functions, and different local coordinates are transformed through each other nonlinearly.

We are considering a massless Dirac field. It is obvious that the introduction of the mass term into the theory does not break the discrete symmetry (\ref{PT_trans_Cont_Minkovski}). Note that the introduction of mass into the Dirac theory on a lattice is paradoxical, since the lattice scale and mass scales of any fermionic fields are incommensurable. According to the prevailing modern concepts, fermion masses arise as a result of certain phase transitions. Since the study of the problem essentially takes place on scales commensurate with Planck's, the masses of fermions can be neglected.

In the theory on a lattice and in the case of the Euclidean signature, the analogue of the PT transformation (\ref{PT_trans_Cont_Minkovski}) is the transformation
\begin{gather}
\left(\Psi^{PT}\right)_{\cV}=U_{PT}\left(\Psi^{\dag}_{\cV}\right)^t, \quad
\left(\Psi^{{\dag}PT}\right)_{\cV}=-\left(\Psi_{\cV}\right)^tU^{-1}_{PT},
\quad  U^{-1}_{PT}\gamma^aU_{PT}=(\gamma^a)^t,
\nonumber \\
\left(\hat{e}^{PT}\right)_{{\cV}_1{\cV}_2}=-U^{-1}_{PT}\hat{e}^t_{{\cV}_1{\cV}_2}U_{PT}, \quad
 (\Omega^{PT})_{{\cV}_1{\cV}_2}=U^{-1}_{PT}\Omega^t_{{\cV}_2{\cV}_1}U_{PT},
\nonumber \\
 U_{PT}=\gamma^1\gamma^3,  \quad  U_{PT}^2=-1.
\label{PT_transform}
\end{gather}
The transformation of Dirac fields in (\ref{PT_transform}) again coincides with the combined transformation of $P$ and $T$ in quantum field theory for the Euclidean signature.
It follows from (\ref{PT_transform}) and (\ref{Dirac_Form}) that \cite{vergeles2021note}
\begin{gather}
\left(\hat{\Theta}^{PT}\right)_{{\cV}_1{\cV}_2}=-U^{-1}_{PT}\hat{\Theta}^t_{{\cV}_1{\cV}_2}U_{PT}.
\label{PT_trans_Dir_Ac}
\end{gather}
Note that in the lattice theory there are no local coordinates at all; therefore, only a variant of the PT transformation (\ref{PT_transform})  is possible.

Since both the lattice theory and its long-wavelength limit are symmetric relative to the described PT symmetry, this means that the PT symmetry is exact, consistent with regularization.

We emphasize that discrete symmetry (\ref{PT_trans_Cont_Minkovski}), (\ref{PT_transform}) is not
a discrete subgroup of the gauge group. This can be seen already from the fact that under gauge transformations the field $\Psi$ is transformed through the field $\Psi$. The same is true for the field $\overline{\Psi}$.
However, the PT transformation swaps the fields $\Psi$ and $\overline{\Psi}$, that is, particles and antiparticles.

From the above, the conclusion follows: if  there is a certain state $|\lambda \rangle$,
then there is also a state $|\lambda; PT \rangle$, obtained from the first by means of the $ PT $-transformation.
Let
\begin{gather}
\langle\lambda|e^a_{\mu}(x)|\lambda\rangle={\me}^a_{\mu}(x), \quad
\langle\lambda|\Theta^a_{\mu}(x)|\lambda\rangle=\theta^a_{\mu}(x).
\label{Mean}
\end{gather}
Then, according to (\ref{PT_trans_Cont_Minkovski}) and (\ref{PT_trans_Dir_Ac_Min})
\begin{gather}
\langle\lambda|\left(e^{PT}\right)^a_{\mu}(x)|\lambda\rangle=-\me^a_{\mu}(x), \quad
\langle\lambda|\left(\Theta^{PT}\right)^a_{\mu}(x)|\lambda\rangle=-\theta^a_{\mu}(x).
\label{Mean_PT_var}
\end{gather}
On the other hand
\begin{gather}
\langle\lambda;PT|\left(e^{PT}\right)^a_{\mu}(x)|\lambda;PT\rangle=\me^a_{\mu}(x), \quad
\langle\lambda;PT|\left(\Theta^{PT}\right)^a_{\mu}(x)|\lambda;PT\rangle=\theta^a_{\mu}(x),
\label{Mean_PT_State_var}
\end{gather}
\begin{gather}
\langle\lambda;PT|e^a_{\mu}(x)|\lambda;PT\rangle=-\me^a_{\mu}(x), \quad
\langle\lambda;PT|\Theta^a_{\mu}(x)|\lambda;PT\rangle=-\theta^a_{\mu}(x).
\label{Mean_PT_State}
\end{gather}

Obviously, the sign of $\langle\Theta^a_{\mu}\rangle$ is determined by the filling rule of the Dirac
sea. To clarify the situation, consider the Dirac Hamiltonian on the hypersurface  $x^0=\const$:
\begin{gather}
{\cal H}_{\Psi}=-\frac12\varepsilon_{\alpha\beta\gamma}\int_{x^0=\const}e^0_0\cdot\Theta^{\alpha}\wedge
e^{\beta}\wedge e^{\gamma}.
\label{Dirac_Grav_Simpl_Ham}
\end{gather}
Recall that $\Theta^{\alpha}\equiv\Theta^{\alpha}_i\d x^i$, $e^{\alpha}\equiv e^{\alpha}_i\d x^i$.
From (\ref{Dirac_Grav_Simpl_Ham}) it is obvious that the reflection
$e\longrightarrow-e$ entails the need to redefinition the Dirac ground state by mutual replacement
particles and antiparticles. The PT transformation (\ref{PT_trans_Cont_Minkovski}) leaves
the Hamiltonian (\ref{Dirac_Grav_Simpl_Ham})  invariant.

In the case $e^{\mu}_a\longrightarrow\delta^{\mu}_a$
\begin{gather}
{\cal H}_{\Psi}=-\int\d^{(3)}x\sum_{\alpha=1}^3\Theta^{\alpha}_{\alpha}
=\int\d^{(3)}x\,\Psi^{\dag}\left(-i\gamma^0\gamma^{\alpha}\partial_{\alpha}\right)\Psi.
\label{Minkovski_Hamiltonian}
\end{gather}
The spectrum of the operator $\left(-i\gamma^0\gamma^{\alpha}\partial_{\alpha}\right)$ has both positive
$|{\bf k}|$ and negative $(-|{\bf k}|)$ eigenvalues, and there is
one-to-one correspondence between positive and negative frequency wave functions.
In the ground state of the system, all levels with negative energy of this operator are filled (Dirac sea), so that all excitations of the Hamiltonian (\ref{Minkovski_Hamiltonian}) turn out to be positive-frequency.
Let us denote the corresponding ground state as $|0\rangle$. Obviously
\begin{gather}
\langle 0|\Theta^{\alpha}_{\alpha}|0\rangle=2\int\frac{\d^{(3)}k}{(2\pi)^3}|{\bf k}|\longleftrightarrow
\langle 0|{\cal H}_{\Psi}|0\rangle=-2V\int\frac{\d^{(3)}k}{(2\pi)^3}|{\bf k}|.
\label{Mean_Vac}
\end{gather}
The $PT$-transformation swaps the positions of the positive and negative frequency wave functions.
This can be seen from the chain of equations, in which each next equation follows from the previous one:
\begin{gather}
-i\gamma^0\gamma^{\alpha}\partial_{\alpha}\psi_N=\epsilon_N\psi_N\longrightarrow
i\partial_{\alpha}\overline{\psi}_N\gamma^{\alpha}\gamma^0=\epsilon_N\overline{\psi}_N
\longrightarrow-i\gamma^0\gamma^{\alpha}\partial_{\alpha}\psi_N^{PT}=(-\epsilon_N)\psi_N^{PT}.
\label{Dirac_Mode_Trans}
\end{gather}
Therefore, the state $|0;PT\rangle$ is constructed by filling all levels with positive energy of the
operator $\left(-i\gamma^0\gamma^{\alpha}\partial_{\alpha}\right)$. We also have
$(\Theta^{PT})^a_{\mu}=-\Theta^a_{\mu}$, $(e^{PT})^{\mu}_a=-\delta^{\mu}_a$, so that
\begin{gather}
{\cal H}^{PT}_{\Psi}=\int\d^{(3)}x\,(\Psi^{PT})^{\dag}\left(-i\gamma^0\gamma^{\alpha}\partial_{\alpha}\right)\Psi^{PT}.
\label{Hamiltonian_PT}
\end{gather}
According to (\ref{Dirac_Mode_Trans}) and (\ref{Hamiltonian_PT}) we again have
\begin{gather}
\langle 0;PT|{\cal H}_{\Psi}^{PT}|0;PT\rangle=-2V\int\frac{\d^{(3)}k}{(2\pi)^3}|{\bf k}|,
\label{Mean_Vac_Complete}
\end{gather}

We will call the state  $|0\rangle$ with a filled Dirac sea as the Dirac vacuum, and the state $|0;PT\rangle$  with a filled anti-Dirac sea (from the point of view of the Dirac vacuum) as the anti-Dirac vacuum.

Next, we need to take into account the fact that the considered field theory is local.
Comparing this fact with the fact that the states $|\lambda\rangle$ and $|\lambda;PT\rangle$ are "equal"
(in the sense that they are translated into each other by a discrete symmetry transformation), naturally to assume that the Universe contains domains with Dirac and anti-Dirac seas.
The domain wall connecting such domains is the subject of this work.

In what follows, the domain with the Dirac vacuum is called domain I, and its adjacent domain with the anti-Dirac vacuum is called domain II.

In quantum field theory, a vast literature is devoted to domain walls.
However, the nature of the studied domains and domain walls was different in the theory of gravity and cosmology: the domains differed in the average values of the Higgs field, but not in the filling of the Dirac sea.
The author is unaware of any papers on domain walls in the theory of gravity that affect the structure of the Dirac sea.
Here we cite just a few foundational works on this topic \cite{zeldovich1974zh,linde1979phase,vilenkin1981gravitational}.
The closest to our work in physical sense (but in condensed matter) are the works
\cite{salomaa1988cosmiclike,volovik1990superfluid,volovik1999vierbein,volovik2019two}.
In these works, the problem is studied in the same variables that are used in the theory of gravity.

The quantum theory of gravity is a non-renormalizable theory. In particular, this means that quantum fluctuations of gravitational dynamical variables (tetrads and connections) are large on ultra-small scales of the order of the Planck length. But on scales much larger than the Planck scale, these fluctuations decay rapidly (according to a power law). Therefore, we will assume that when considering physics on scales that are much larger than Planck's, fluctuations of the gravitational degrees of freedom are insignificant, that is, these degrees of freedom are described by classical fields. In this case, the quantum fluctuations of the Dirac fields are taken into account by averaging over the Dirac vacuum of all operators constructed using the Dirac fields.
Thus, Dirac propagators calculated in an external classical gravitational field are introduced into the computational procedure.

Let the normal Riemann coordinates $x^{\mu}$ be introduced near some point, so that the point $p$ is their center
and $x^{\mu}(p)=0$. Since we do not consider interactions other than gravitational here, near the point $p$
\begin{gather}
\langle0|\hat{T}\Psi(x)\overline{\Psi}(y)|0\rangle=iS_c(x,y)=\left(i\gamma^{\nu}\partial_{\nu}\right)
\frac{1}{4\pi^2\left(-(x-y)^2+l_P^2+i\varepsilon\right)}.
\label{Dir_Propagator}
\end{gather}
On the right side (\ref{Dir_Propagator}), the denominator of the standard propagator $\left(-(x-y)^2+i\varepsilon\right)$ is replaced by $\left(-(x-y)^2+l_P^2+i\varepsilon\right)$.
This change simulates a Dirac sea of finite depth, which takes place in the case of lattice regularization.
It is easy to check that in (\ref{Dir_Propagator}) the positive constant $l_P^2$ is equal (in order of magnitude) to the square of the Planck or lattice scale. Indeed, if $(x^0-y^0)\longrightarrow-0$, then
\begin{gather}
\langle0|\overline{\Psi}_{\beta}(y)\Psi_{\alpha}(x)|0\rangle=-iS_c(x,y)_{\alpha\beta},
\label{Mean_overline(Psi)_Psi}
\end{gather}
and the energy density of the Dirac vacuum
\begin{gather}
\langle0|\overline{\Psi}(x)(-i\gamma^{\alpha}\partial_{\alpha})\Psi(x)|0\rangle=
\tr\left[(-i\gamma^{\alpha}\partial_{\alpha})(-iS_c(x,y))\right]\big|_{x-y=-0}=
-\frac{6}{\pi^2l_P^4}.
\label{Dirac_vac_energy}
\end{gather}
On the other hand, direct calculation of the energy density of the Dirac vacuum leads to the same answer:
\begin{gather}
-2\int\frac{\d^{(3)}k}{(2\pi)^3}k\cdot e^{-l_Pk}=-\frac{6}{\pi^2l_P^4}.
\label{Direct_Dirac_vac_energy}
\end{gather}
Comparison of the right-hand sides of the last two equations implies the above statement.
Note that in our theory we also have in the formula (\ref{Dir_Propagator}) $\varepsilon\sim l^2_P$.

Now we can calculate the vacuum mean of the operator $\Theta^a_{\mu}$ in normal Riemann coordinates near the point $p$:
\begin{gather}
\langle0|\Theta^a_{\mu}|0\rangle|_p=i\tr\left[\gamma^a\partial_{\mu}(-iS_c(x,y))\right]\big|_{x-y=0}=
\frac{2}{\pi^2l^4_P}\delta^a_{\mu}.
\label{Theta_Mean}
\end{gather}
Hence, it is obvious that in arbitrary coordinates the equality
\begin{gather}
\langle0|\Theta^a_{\mu}(x)|0\rangle=f(x)e^a_{\mu}(x),
\label{Theta_Mean_forWall}
\end{gather}
where $f(x)$ is some scalar function. The fact that the relation $e^a_{\mu}\propto\Theta^a_{\mu}$ takes place in the theory under consideration was previously contained in the works
\cite{volovik2021dimensionless,diakonov2011towards,vladimirov2012phase,vladimirov2014diffeomorphism}.

In our case, the Einstein equation is convenient to use in the form
\begin{gather}
\delta {\mA}/\delta e^a_{\mu}=0.
\label{Einstein_Eq}
\end{gather}
There is also an equation obtained by varying the action with respect to the connection $\omega^{ab}_{\mu}$:
\begin{gather}
\delta {\mA}/\delta \omega^{ab}_{\mu}=0  \longrightarrow
\nabla_{\mu}e^a_{\nu}-\nabla_{\nu}e^a_{\mu}=
\frac{il_P^2}{4}e^b_{\mu}e^c_{\nu}\overline{\Psi}\left(\gamma^a\sigma_{bc}+\sigma_{bc}\gamma^a\right)\Psi=0.
\label{Dirac_Torsion}
\end{gather}
The last equality follows from the fact that, according to (\ref{Dir_Propagator}), the Dirac field propagator is linear in the $\gamma$-matrices, and also from the identity
\begin{gather}
\tr\left[\gamma^d\left(\gamma^a\sigma^{bc}+\sigma^{bc}\gamma^a\right)\right]\equiv0.
\label{tr_gamma^4_identity}
\end{gather}

Let's make an infinitesimal transformation of coordinates $x^{\mu}\longrightarrow x^{\mu}-\xi^{\mu}(x)$. As known, in this case, the metric tensor and the tetrad change according to the formulas
\begin{gather}
\delta g_{\mu\nu}=\nabla_{\mu}\xi_{\nu}+\nabla_{\nu}\xi_{\mu}, \quad
\delta e^a_{\mu}=\nabla_{\mu}\xi^a, \quad \xi^a=e^a_{\mu}\xi^{\mu}.
\label{Tetrad_Variation}
\end{gather}
The variation of the Dirac action (\ref{Long_Wav_Dir_Act}) on the mass shell with respect to the variation (\ref{Tetrad_Variation}) is equal to zero, and the direct calculation gives
\begin{gather}
\delta\mA_{\Psi}=
\frac12\varepsilon_{abcd}\varepsilon^{\mu\nu\lambda\rho}\int
\langle0|\Theta^a_{\mu}|0\rangle\,e^b_{\nu}\,e^c_{\lambda}\,\nabla_{\rho}\xi^d
\d x^0\wedge\d x^1\wedge\d x^2\wedge\d x^3
\nonumber \\
=\frac12\varepsilon_{abcd}\varepsilon^{\mu\nu\lambda\rho}\int
fe^a_{\mu}\,e^b_{\nu}\,e^c_{\lambda}\,\nabla_{\rho}\xi^d
\d x^0\wedge\d x^1\wedge\d x^2\wedge\d x^3=0.
\label{Var_Dir_Action}
\end{gather}
Here we have used the equalities (\ref{Theta_Mean_forWall}) and (\ref{Tetrad_Variation}).
Since the right side of Equation (\ref{Dirac_Torsion}) is equal to zero, then the right-hand side (\ref{Var_Dir_Action}) is rewritten as
\begin{gather}
\delta\mA_{\Psi}=
-\frac12\varepsilon_{abcd}\varepsilon^{\mu\nu\lambda\rho}\int\xi^d\left(\partial_{\rho}f\right)
e^a_{\mu}\,e^b_{\nu}\,e^c_{\lambda}\d x^0\wedge\d x^1\wedge\d x^2\wedge\d x^3=0,
\label{Var_2_Dir_Action}
\end{gather}
whence follows the equation
\begin{gather}
\varepsilon_{abcd}\varepsilon^{\mu\nu\lambda\rho}e^a_{\mu}\,e^b_{\nu}\,e^c_{\lambda}\partial_{\rho}f=0.
\label{Eq_Motion_Dir}
\end{gather}
The equation (\ref{Eq_Motion_Dir}) is in our case the equation of motion of matter, which in the general case has the form $\nabla_{\mu}T^{\mu\nu}=0$, where $T^{\mu\nu}$ is the energy-momentum tensor of matter.

Thus, to study the domain wall, we must solve the system of equations (\ref{Einstein_Eq}), (\ref{Dirac_Torsion}) and (\ref{Eq_Motion_Dir}) for the classical fields $ e^a_{\mu}$, $\omega^{ab}_{\mu}$ and $f$.

\section{domain wall}

We denote the coordinates on the space-like hyperplane $\Sigma$ by $(x,\,y,\,z)$.
Suppose that the domain wall is flat and it
is parallel to the plane $ z = 0 $ in the hyperplane $\Sigma$.

It follows from the assumptions made that the metric tensor depends only on the coordinates $(t,\,z)$ and is invariant relative to rotations around any perpendicular to the plane $(x,\,y)$. This means that all off-diagonal elements of the metric tensor, except for $g_{t\,z}$, are equal to zero. Then
\begin{gather}
g_{\mu\nu}=\big(g_{tt},\,g_{xx},\,g_{yy},\,
g_{zz},\,g_{tz}\big)=\big(\rho^2(t,z),\,-\lambda^2(t,z),\,-\lambda^2(t,z),\,-\sigma^2(t,z),\,\tau(t,z)\big).
\label{Metrics}
\end{gather}
From the form of the metric (\ref{Metrics}) it follows that the tetrad can be chosen so that
$e^{\alpha}_i\sim\delta^{\alpha}_i$ for $\alpha= 1,2$, $i=1,2$.
In this case, two of the four equations (\ref{Eq_Motion_Dir})
with $d=1,2$ are satisfied identically. The other two equations with $d= 0,3$ are reduced to the form
\begin{gather}
\left(e^3_3\partial_t-e^3_0\partial_z\right)f=0,  \quad
\left(e^0_3\partial_t-e^0_0\partial_z\right)f=0.
\label{Eq_Motion_2_Dir}
\end{gather}

A pair of equations (\ref{Eq_Motion_2_Dir}) has a nonzero solution only if
\begin{gather}
e^0_0e^3_3-e^0_3e^3_0=0.
\label{Self-consistency}
\end{gather}
The last relation means that the metric is degenerate near the domain wall.

Let's consider several options for solutions.

\subsection{Stationary solution}

A stationary solution to the system (\ref{Eq_Motion_2_Dir}), when $f(z)$ depends on $z$, but does not depend on $t$, is physically meaningless. Indeed, in this case, for the fulfillment of the equations
(\ref{Eq_Motion_2_Dir}) requires $e^0_0=e^3_0=0$, and the metric takes the form
\begin{gather}
\d s^2=\left[\left(e^0_3\right)^2-\left(e^3_3\right)^2\right]\d z^2-\lambda^2\left(\d x^2+\d y^2\right).
\label{Wrong}
\end{gather}
Thus, if the square bracket in (\ref{Wrong}) is less than zero, then the interval is always less than or equal to zero, and there is no proper time. Otherwise, the $z$ coordinate becomes the time coordinate, which is also
meaningless.

\subsection{The case $f=\const$}

Let's consider the simplest solution
\begin{gather}
\partial_{\rho}f=0 \longrightarrow f=\const,
\label{Const}
\end{gather}
satisfying equations (\ref{Eq_Motion_Dir}). In this case, for the existence a physically acceptable solution requires the condition
\begin{gather}
\tilde{f}\equiv f+\Lambda=0.
\label{Condition_Planar}
\end{gather}
Indeed, from (\ref{Condition_Planar}) it follows that only the term contributes to the Einstein equation (\ref {Einstein_Eq}) $\mA_g$. According to (\ref{Long_Wav_Grav_Act}), all components $\delta{\mA}_g/\delta e^a_{\mu}$ are proportional to curvature tensor $\mR^{ab}$, which is bilinear with respect to the first derivatives and linear with respect to the second derivatives of the tetrad. For a physically acceptable solution, all derived tetrads must vanish far from the wall, that is, far from the wall, the equality $\delta{\mA}_g/\delta e^a_{\mu}=0$ is necessary. If we assume that the cosmological constant can only change on scales,
significantly larger scales of the domain wall, then the condition (\ref{Condition_Planar}) must be satisfied at all points in space-time. Thus, in this case, the Einstein equation is reduced to the Einstein equation in emptiness.

Let us write down the set of Einstein's equations.

It is natural to assume that in the case (\ref{Const}) the metric is nondegenerate. Then, by transforming the coordinates $(t,\, z)\longrightarrow (t',\,z')$, one can vanish the only nonzero off-diagonal component of the metric $g_{tz}$. As a result, the tetrad can be chosen in the form
\begin{gather}
e^0=\rho\d t, \quad e^1=\lambda\d x, \quad  e^2=\lambda\d y, \quad  e^3=\sigma\d z.
\label{Tetrad}
\end{gather}
We will also assume that all geometric quantities depend only on $z$. Using equations
(\ref{Dirac_Torsion}) find all components of the connection coefficients and then all components of the Riemann curvature tensor $\mR^{ab}_{\mu\nu}$, whose nonzero components
\begin{gather}
\mR^{01}_{01}=\mR^{02}_{02}=\frac{\rho'\lambda'}{\sigma^2}, \quad
\mR^{03}_{03}=\left(\frac{\rho''}{\sigma}-\frac{\rho'\sigma'}{\sigma^2}\right),
\nonumber \\
\mR^{12}_{12}=\frac{\lambda'^2}{\sigma^2},  \quad
\mR^{13}_{13}=\mR^{23}_{23}=\left(\frac{\lambda''}{\sigma}-\frac{\lambda'\sigma'}{\sigma^2}\right).
\label{Riemann_tensor}
\end{gather}
Using the equations (\ref{Tetrad}) and (\ref {Riemann_tensor}), we write out all independent and not identically zero Einstein's equations (\ref{Einstein_Eq}) in the absence of matter:
\begin{gather}
\lambda'^2+2\lambda\lambda''-2\frac{\lambda\lambda'\sigma'}{\sigma}=0,
\nonumber \\
\rho\lambda''+\rho''\lambda+\rho'\lambda'-\frac{\sigma'}{\sigma}\left(\rho\lambda\right)'=0,
\nonumber \\
2\rho'\lambda+\rho\lambda'=0.
\label{Einstein_Eq_Vac}
\end{gather}
The last equation has the integral $\rho^2\lambda=\const$. Making the substitution $\rho=\const\cdot\lambda^{-1/2}$ in the second equation, we arrive at the first of the equations
(\ref{Einstein_Eq_Vac}) which is easy to integrate. Thus, we find:
\begin{gather}
\lambda=\rho^{-2}, \quad \sigma=A\cdot\sqrt{\lambda}\lambda'.
\label{Einstein_Eq_Vac_2}
\end{gather}
Here $A$ is some constant. Two equations (\ref{Einstein_Eq_Vac_2}) for three functions admit many solutions, the selection of which is not clear.

Let domain I be located in the region $z>0$. Consider, for example, the following solution in domain I:
\begin{gather}
\rho=\frac{(a^2+z^2)^{1/4}}{\sqrt{z}}, \quad  \lambda=\frac{z}{\sqrt{a^2+z^2}},  \quad
\sigma=A\frac{\sqrt{z}}{(a^2+z^2)^{7/4}}, \quad  a=\const.
\label{Solution_2}
\end{gather}
Obviously, in the region $z<0$, that is, in domain II we have the following solution to the system of equations
(\ref{Einstein_Eq_Vac}):
\begin{gather}
-\rho(-z), \quad -\lambda(-z), \quad -\sigma(-z).
\label{Z_plane_reflection}
\end{gather}
Thus, a tetrad is constructed that satisfies the Einstein equation everywhere except for the plane $z=0$ and which is antisymmetric with respect to this plane. According to (\ref{Theta_Mean_forWall}), together with the tetrad, the average $\Theta^a_{\mu} $ changes its sign. In this case, the time part of the tetrad $\rho$ jumps through infinity.

From the above formulas it follows that for a given solution the tetrad in the region $z<0$ cannot be obtained from the tetrad in the region $z>0$ by analytic continuation. If we take into account that we assume a lattice regularization of the theory, then this fact does not seem to be inadmissible.

We point out that the total action of the system remains finite near the domain wall. Indeed, we used the equations $\delta{\mA}_g/\delta e^a_{\mu}=0$, from which it follows that the Ricci tensor is equal to zero. In this case, the gravitational part of the action also vanishes. We also have:
\begin{gather}
\mA_{\Psi}+\mA_{\Lambda}\sim e^0_0e^1_1e^2_2e^3_3=\rho\lambda^2\sigma\sim \frac{z^2}{(a^2+z^2)^{5/2}}.
\nonumber
\end{gather}
This quantity converges upon integration along the $z$ axis. This implies the finiteness of the total action of the system for a unit of 4-volume including the final section of the domain wall.

As mentioned above, the author is unaware of any papers on domain walls in the theory of gravity that affect the structure of the Dirac sea. Therefore, generally speaking, there can be no direct comparison of our results with the results of other works on domain walls. Nevertheless, such a comparison can be made in this Subsection, since the condition (\ref{Condition_Planar}) actually excludes the Dirac sea from consideration.
Thus, equations (\ref{Condition_Planar})-(\ref{Einstein_Eq_Vac_2}) also have a Kasner solution in domain I
(see \cite{vilenkin1981gravitational}, Eq. (36)):
\begin{gather}
A=\frac32, \quad \rho=z^{-1/3},  \quad \lambda=z^{2/3}, \quad \sigma=1.
\label{Kasner}
\end{gather}
Apparently, this solution is even less suitable for describing the domain wall we need.

\subsection{Time-dependent solution}

Let us consider the case of a domain wall depending on one spatial coordinate and on time.
From a physical point of view, it seems reasonable to consider the case when the domain wall moves parallel to itself at a constant speed. Obviously, the meaningful case takes place if the wall moves along the $z$ axis with the speed of light. Otherwise, the problem is reduced using the Lorentz transformation to the problem already considered in Section III A.
In other words, all variables must depend on the combination $\psi=(t\pm z)$. Consider the possibility $\psi=(t-z)$. Then the equations (\ref{Eq_Motion_2_Dir}) take the form
\begin{gather}
e^0_0+e^0_3=0,  \quad  e^3_0+e^3_3=0.
\label{Eq_Motion_3_Dir}
\end{gather}
Thus
\begin{gather}
e^0=\rho(\d t-\d z), \quad  e^1=\lambda\d x,  \quad  e^2=\lambda\d y, \quad e^3=\sigma(\d t-\d z).
\label{Tetrad_3}
\end{gather}
We assume that $|\sigma|<|\rho|$. The opposite case has no physical meaning, since then the metric would not have timelike intervals. With the help of the (local) Lorentz transformation, we vanish the component of the tetrad $e^3$, which is possible due to the inequality $|\sigma|<|\rho|$. Although the tetrad (\ref{Tetrad_3}) is degenerate, it is easy to prove that in the used gauge the equation (\ref{Dirac_Torsion}) has only one solution for the connection. Indeed, in the gauge $e^3=0$ we have $\omega_{3a}=-\omega_{a3}=0$. Obviously, in the expansion $\omega_{ab}=\gamma_{abc}e^c$ we can assume that $\gamma_{ab3}=0$. Thus, $\gamma_{abc}=0$ if at least one of the indices $a,b,c$ is 3. The remaining components $\gamma_{abc}$ with indices $a,b,c=0,1,2$ are determined unambiguously in the standard way using equations (\ref{Dirac_Torsion}).

Bypassing intermediate calculations, we write out the nonzero components of the Riemann tensor $\mR^{ab}_{\mu\nu}$:
\begin{gather}
\mR^{01}_{01}=\mR^{01}_{13}=\mR^{02}_{02}=\mR^{02}_{23}=-\frac{\lambda''}{\rho}+\frac{\lambda'\rho'}{\rho^2},
\quad \mR^{12}_{12}=-\frac{\lambda'^2}{\rho^2}.
\label{Riemann_tensor_3}
\end{gather}
In this subsection, everywhere $F'$ means the derivative of the function $F$ with respect to the variable $\psi=(t-z)$. All components of the Einstein equation  are reduced to one equation:
\begin{gather}
-\frac{\lambda\lambda''}{\rho}-\frac{\lambda'^2}{2\rho}+\frac{\lambda\lambda'\rho'}{\rho^2}=\frac34
l_P^2\tilde{f}\rho\lambda^2,
\nonumber \\
\tilde{f}(\psi)=\tilde{f}(-\psi),
\quad |\tilde{f}|\longrightarrow 0 \quad \mbox{as} \quad |\psi|\longrightarrow\infty.
\label{Einstein_Eq_3}
\end{gather}
We assume that in domain I the variable $\psi>0$, and in domain II $\psi<0$.

As above, Einstein's equation does not fix the metric. Consider the following solution to the equation (\ref{Einstein_Eq_3}) in the region $\psi>0$:
\begin{gather}
\rho=\frac{\sqrt{\psi}}{(a^2+\psi^2)^{1/4}}, \quad  \lambda=\frac{\psi}{\sqrt{a^2+\psi^2}},  \quad
a=\const\gg l^2_P.
\label{Solution_3}
\end{gather}
The solution in the region $\psi<0$ is obtained from the solution (\ref{Solution_3}) by reflection relative to the plane $\psi=0$ according to (\ref{Z_plane_reflection}) with the replacement of $z\longrightarrow\psi$. From equations (\ref{Einstein_Eq_3}) and (\ref{Solution_3}) we find:
\begin{gather}
\tilde{f}=\frac{4a^2}{l^2_P|\psi|(a^2+\psi^2)^{3/2}}.
\label{Theta}
\end{gather}

\section{Discussion}

Option C seems to be the most interesting. In coordinates $(t,\,x,\,y,\,z)$ metric has the form
\begin{gather}
\d s^2=\rho^2(\d t-\d z)^2-\lambda^2\left(\d x^2+\d y^2\right).
\label{Wall_Metric}
\end{gather}
This metric is degenerate.

We assume that in the region
\begin{gather}
\psi^2\gg a^2\gg l^2_P
\label{Away_from_Wall}
\end{gather}
the metric approaches non-degenerate and relaxes to its normal (quasi) flat form.

Consider a constant time surface $t=t_0$, $\d t=0$, $\d y=0$. The variable $\psi$ becomes dependent only on the coordinate $z$: $\psi=\pm(z-t_0)$. Then, according to (\ref{Solution_3}) and (\ref{Wall_Metric}), in the case of $\psi\longrightarrow0$ the solution to the equation $\d s=0$ has the form
\begin{gather}
z-t_0=\pm\frac{1}{4a}(x-x_0)^2.
\label{Wall_Approach}
\end{gather}
Thus, upon displacement near the domain wall along the curve
(\ref{Wall_Approach}) we have $\int\d s=0$, that is, the physical length of such an arc is zero.
It follows from this that in a model description of the domain wall, it can be considered infinitely thin.
But in this case it does not matter which solution of the equation
(\ref{Einstein_Eq_3}) is chosen because only the asymptotics for $|\psi|\longrightarrow\infty$ are important.

Since the metric is degenerate near the domain wall, it is obvious that near the domain wall the action tends to zero, that is, the action is finite.

The text below is not a complete scientific result, but material for research, since some important aspects remain unresolved here. In particular, the author considers it necessary to thoroughly study the formulas (\ref{Number_of_Particles})-(\ref{Mean_Number_of_Particles}) and the physics described by these formulas.

Consider a conserved Dirac current $J^a=\overline{\Psi}\gamma^a\Psi$ near the domain wall. To calculate it, we use the general formula
\begin{gather}
J^a=\Re\Big\{-i\tr\gamma^aS_c(x,y)|_{|{\bf x}-{\bf y}|=0,\,x^0-y^0\longrightarrow-0}\Big\}.
\label{Ferm_Current}
\end{gather}
Hence, using the formulas (\ref{Dir_Propagator}) and (\ref{Mean_overline(Psi)_Psi}) and the fact that here $\varepsilon\sim l^2_P$, we find
\begin{gather}
J^a=-\frac{1}{\pi^2}\eta^{a\nu}\partial_{\nu}\frac{\varepsilon}{\xi^2+\varepsilon^2}\Big|_{|{\bf x}-{\bf y}|=0,\,x^0-y^0=-l_P}, \quad \xi=-(x-y)^2+l^2_P.
\label{Ferm_Current_Wall}
\end{gather}
Obviously, in (\ref{Ferm_Current_Wall}) we can put $\xi=0 $ even before differentiation, and therefore
\begin{gather}
J^a(x)=-\frac{1}{\pi^2}\eta^{a\nu}\partial_{\nu}\frac{1}{\varepsilon}\sim
-\frac{1}{\pi^2}\eta^{a\nu}\partial_{\nu}\frac{1}{l^2_P(\psi).}
\label{Ferm_Current_Wall_Fin}
\end{gather}
According to (\ref{Theta_Mean}) and (\ref{Theta_Mean_forWall}) $f=2/(\pi^2l_P^4)$. Therefore, for the current
(\ref{Ferm_Current_Wall_Fin}) we get:
\begin{gather}
J^a=-\frac{1}{4}l_P^2\eta^{a\nu}\partial_{\nu}f=-\frac{1}{4}l_P^2\eta^{a\nu}\partial_{\nu}\tilde{f}.
\label{Ferm_Current_Wall_2}
\end{gather}
Substituting here $\tilde{f}$ from (\ref{Theta}), we finally find the formula for the charge density of fermions
in domain I:
\begin{gather}
J^0(\psi)= \frac{a^2(a^2+4\psi^2)}{|\psi|^2(a^2+\psi^2)^{5/2}}>0.
\label{Particle_Density}
\end{gather}
According to (\ref{Particle_Density}) $J^0>0$, that is, near the domain wall, the density of particles (but not antiparticles) is finite.

The charge (number of particles)
\begin{gather}
{\cal N}=\frac16\varepsilon_{abcd}\int_{\Sigma}J^ae^b\wedge e^c\wedge e^d
\label{Number_of_Particles}
\end{gather}
is conserved if $\nabla_aJ^a=0$. The latter property takes place in the theory under consideration.
Here $\Sigma$ there is a certain  space-like hypersurface.
Note that the lattice analog of the number of particles is also strictly conserved. According to
(\ref{PT_trans_Cont_Minkovski}) $(J^{PT})^a=J^a$. By definition, in domain I, the ground state is denoted
$|0\rangle$, and in domain II, as $|0;PT\rangle$. Thus, we have:
\begin{gather}
\langle0|J^a|0\rangle=\langle0|(J^{PT})^a|0\rangle=\langle0;PT|J^a|0;PT\rangle.
\label{Denc_PT_Trans}
\end{gather}
We split the integral (\ref{Number_of_Particles}) into 2 integrals: the integral over domain I and over domain II and denote the corresponding integrals as ${\cal N}_I$ and ${\cal N}_{II}$. Then the average of the number of particles (\ref{Number_of_Particles}) is divided into 2 terms:
\begin{gather}
\langle{\cal N}\rangle=\langle{\cal N}_I\rangle+\langle PT|{\cal N}_{II}|PT\rangle=\const.
\label{Mean_Number_of_Particles}
\end{gather}
According to (\ref{Particle_Density}) we have $\langle{\cal N}_I\rangle>0$, while according to
(\ref{Mean_PT_State}) and (\ref{Denc_PT_Trans}) $\langle PT|{\cal N}_{II}|PT\rangle<0$.
In this case, the sum (\ref{Mean_Number_of_Particles}) can be equal to zero, which is further assumed.

Consider the particle density in the region $\psi>\psi_0=t_0-z_0\gg a$. Let us introduce the particle distribution function $\delta n(\psi,{\bf p})$, so
\begin{gather}
J^0(\psi)=\int\frac{\d^{(3)}p}{(2\pi)^3}\delta n(\psi,{\bf p}).
\label{Distribution_Function}
\end{gather}
It is very important here that the quasi-momentum of a quasiparticle on an irregular "breathing" lattice
\cite{vergeles2021note} cannot be strictly conserved. Therefore, it is natural to assume that $\delta n(\psi,{\bf p})>0$ for any ${\bf p}$.
Let us write out the kinetic equation for this distribution function:
\begin{gather}
\frac{\partial\delta n}{\partial t}+\frac{\partial\epsilon}{\partial{\bf p}}\cdot \frac{\partial\delta n}{\partial{\bf r}}-
\frac{\partial\epsilon}{\partial{\bf r}}\cdot \frac{\partial\delta n}{\partial{\bf p}}=0.
\label{Kinetic_Equation}
\end{gather}
We assume that $\epsilon({\bf p})=|{\bf p}|$, and omit the collision integral on the right-hand side
of kinetic equation. This means that we neglect dissipative phenomena in further reasoning, which does not devalue the qualitative conclusions. Since $\partial\epsilon/\partial{\bf r}=0$, the last term in the equation (\ref{Kinetic_Equation}) is missing.

We have $\partial\epsilon/\partial{\bf p}={\bf p}/\epsilon={\bf n}$. For the momentum ${\bf p}_-=(0,0,-|{\bf p}|)$, directed in the opposite direction from the domain wall, equation (\ref{Kinetic_Equation})
takes the form
\begin{gather}
\frac{\partial\delta n}{\partial t}-\frac{\partial\delta n}{\partial z}=0.
\label{Kinetic_Equation(-)}
\end{gather}
This equation can be easily integrated: on the trajectory $t(s)=s+t_0$, $z(s)=-s+z_0$, the function $\delta n$ remains constant:
\begin{gather}
\delta n(t=s+t_0,\,z=-s+z_0,\,{\bf p}_-)=\const>0.
\label{Finite_Posit_dens-}
\end{gather}
On the same trajectory $\psi(s)=2s+\psi_0$. Of course the $s$-parameter in the solution (\ref{Finite_Posit_dens-}) should remain small enough, since the collision integral was not taken into account when finding the solution.

Equation (\ref{Finite_Posit_dens-}) shows that there is a dynamic tendency to the emergence of undamped positive particle density at any distance from the domain wall. It was said above that the total number of particles
(see Eq. (\ref{Mean_Number_of_Particles})) is conserved. The question of where antiparticles accumulate remains open.

Although the presentation in this Section is very schematic, nevertheless it seems to us that the question "is not gravity a source of baryon asymmetry in the Universe?" deserves consideration.

\begin{acknowledgments}

I am grateful to G. Volovik for stimulating attention to this work.
This work was carried out as a part of the State Program 0033-2019-0005.

\end{acknowledgments}


\end{document}